\documentclass[prl,twocolumn,aps]{revtex4}
\usepackage{graphicx}
\usepackage{amsfonts}
\begin{document}

\newcommand{\be}{\begin{equation}}
\newcommand{\ee}{\end{equation}}
\newcommand{\bea}{\begin{eqnarray}}
\newcommand{\eea}{\end{eqnarray}}
\newcommand{\beq}{\begin{equation}}
\newcommand{\eeq}{\end{equation}}
\newcommand{\beqn}{\begin{eqnarray}}

\newcommand{\eeqn}{\end{eqnarray}}
\newcommand{\ack}[1]{{\bf Pfft! #1}}
\newcommand{\pa}{\partial}
\newcommand{\osigma}{\overline{\sigma}}
\newcommand{\orho}{\overline{\rho}}
\newcommand{\myfig}[3]{
	\begin{figure}[ht]
	\centering
	\includegraphics[width=#2cm]{#1}\caption{#3}\label{fig:#1}
	\end{figure}
	}
\newcommand{\littlefig}[2]{
	\includegraphics[width=#2cm]{#1}}
\newcommand{\1}{{\rm 1\hspace*{-0.4ex}%
\rule{0.1ex}{1.52ex}\hspace*{0.2ex}}}

\title{
Gauge Fields, Membranes and Subdeterminant Vector Models}
\author{Robert G. Leigh}
\affiliation{Department of Physics, University of Illinois, 1110 W. Green Street, Urbana IL 61801, U.S.A.}
\author{Andrea Mauri}
\affiliation{Dipartimento di Fisica Teorica, Universita degli Studi di Milano, Via Celoria, Milano 20133, Italy}
\author{Djordje Minic}
\affiliation{IPNAS, Department of Physics, Virginia Tech, Blacksburg, VA 24061, U.S.A.} 
\author{Anastasios C. Petkou}
\affiliation{Department of Physics, University of Crete, GR-71003, Heraklion, Greece}


\date{\today}
\begin{abstract}
We present a class of classically marginal $N$-vector models in $d=4$ and $d=3$ whose scalar potentials can be written as subdeterminants of symmetric matrices.  The $d=3$ case is a generalization of the scalar Bagger-Lambert-Gustavsson (BLG) model. Using the Hubbard-Stratonovich transformation we calculate their effective potentials which exhibit intriguing 
large-$N$ scaling behaviors.  We comment on the relevance of our models to strings, membranes and also to a class of novel spin systems that are based on ternary commutation relations.

\end{abstract}

\maketitle


The relationship between $D3$-branes and 4d Yang-Mills theories is a fundamental ingredient in our current understanding of string theory. One of  the simplest manifestations of such a relationship arises in the study of the scalar part  of the action of ${\cal N}=4$ $U(N)$ super-Yang-Mills (SYM) 
\beq
\label{D3action}
I =\int d^4x\ {\rm Tr}\left(\frac{1}{2}\partial_\mu\Phi^I\partial_\mu\Phi^I -\frac{1}{4}[\Phi^I,\Phi^J]^2\right)
\eeq
where $\Phi^I$ are matrix fields $\Phi^I=\Phi^I_\alpha T^\alpha$, $I=1,2,..,6$ and $T^\alpha$, $\alpha=1,2,.. ,N^2$ are the adjoint generators of $U(N)$. Expanding around the Coulomb vacuum, i.e. the $N$ Cartan generators $\overline{\Phi}^I_a$, $a=1,2,..,N$, of $U(N)$ one obtains an effective potential of the form \cite{Zarembo}
\beq
\label{zarembo}
V_{{\rm eff}} \sim \sum_{a<b}\,m_{ab}^4\ln\frac{m_{ab}^2}{\Lambda^2}\,, a,b=1,2,..N\,,
\eeq
where $m_{ab}^2=|\overline{\Phi}^I_a-\overline{\Phi}^I_b|^2$ and $\Lambda$ is a cutoff. 
The Cartan elements $\overline{\Phi}^I_a$ are interpreted as the positions of the $N$ D3-branes 
and the potential is minimized when the branes form a spherical shell of radius $|\overline{\Phi}^I_a|=R\propto \Lambda$ whose energy density is ${\cal E}\propto N^2\Lambda^4$. This simplified analysis reveals the crucial physical property of the system: the potential energy of $N$ $D3$-branes is due to the two-body interactions among them. The latter are naturally interpreted as strings stretched between pairs of branes, having (masses)$^2$ $m_{ab}^2$. Then, for large-$N$ the usual $N^2$ YM scaling arises from the combinatoric factor counting two-body interactions.

The effective potential (\ref{zarembo}) depends essentially on the $N$-vector fields  $\overline{\Phi}^I_a$, through the symmetric composite quantities $m_{ab}^2$, however, only the $N(N-1)/2$ off-diagonal elements of $m_{ab}$ enter the result (\ref{zarembo}). This effect is a consequence of the underlying $U(N)$ algebraic structure of the system. Such a point of view motivates us to ask whether we could capture the essential physics of the $D3$-brane system using a simpler $N$-vector model. Indeed, using  scalars $\overline{\Phi}^I_a$ we could construct various composite quantities that are symmetric in the Cartan indices and then we could conceive an alternative mechanism leading to an effective potential similar to (\ref{zarembo}) without having to assume detailed  knowledge of the underlying algebraic structure. We will present such a model below.  Of course, the knowledge of the $U(N)$ structure provides us with a wealth of additional information regarding the $D3$-brane system and its relationship with string theory. A corresponding model exists in $d=3$, and we believe that it is a fruitful way of approaching the universality class of the conformal theory describing $N$ $M2$-branes \cite{bl}, as in that case the underlying algebraic structure is still not understood.

To motivate the $d=4$ model, consider the scalar potential of the action (\ref{D3action}) for $SU(2)$ generalized to $I,J=1,2,..,N_f$
\beq
\label{SU2pot}
V_{SU(2)}(\Phi^I_a) = \frac{1}{4}\Phi^I_a\Phi^J_b\Phi^I_e\Phi^J_f\epsilon^{abc}{\epsilon^{ef}}_{c}\,,
\eeq
where $a,b,c=1,2,3$. Generalizing $\epsilon_{abc}$ to the structure constants $f_{abc}$  of $U(N)$ with $a,b,c=1,2,..,N^2$ one obtains the scalar potential of $U(N)$ YM with $N_f$ flavors. There is, however, another generalization of (\ref{SU2pot}) with $a,b,c=1,2,..,N$ which is 
\beq
\label{sd2pot}
V_{N}^{(4)}(\Phi^I_a)= \frac{1}{2(N-2)!}\Phi^I_a\Phi^J_b\Phi^I_e\Phi^J_f\epsilon^{abc\cdots}{\epsilon^{ef}}_{c\cdots}\,.
\eeq
Here, we use the appropriate $N$-index $\epsilon$ tensors with $N-2$ indices contracted. We refer to the classically marginal $N$-vector potential (\ref{sd2pot}) as the {\it 2-subdeterminant potential}, and we will find that it has a natural generalization  to other dimensions. 
A nice property of the $N$-vector model with the potential (\ref{sd2pot}) is that it can be analyzed via a Hubbard-Stratonovich (HS) transformation \cite{fradkin}. 
To do so,  we first note that the potential depends only on the $N\times N$ symmetric matrix field
$\rho_{ab}=\Phi^I_a\Phi^I_b$
in terms of which the potential may be written as the sum of all possible 
2$\times$ 2 subdeterminants  of $\rho$ as
\beq
\label{sd2potrho}
V_{N}^{(4)}(\rho)\equiv \ {\det}_2^{(N)}(\rho)=\frac{1}{2}\left[ (tr\rho)^2-tr\rho^2\right]\,.
\eeq
Now we introduce the {\it 4d subdeterminant model} as
\beq
L_{4d} =\frac{1}{2}\Phi^I_a(-\partial^2)\Phi^I_a +\frac{\mu_0^2}{2}\Phi^I_a\Phi^I_a -g_0V_{N}^{(4)}(\Phi^I_a)\,,
\eeq
where $\mu_0^2$ is a bare mass term and $g_0$ a dimensionless coupling. Rescaling then $\Phi^I_a\mapsto \Phi^I_a/\sqrt{g_0}$ and introducing the HS fields $\sigma_{ab}$ and $\rho_{ab}$ we write the partition function as
\beqn\label{D3H-S}
{\cal Z}&=&\int [{\cal D}\Phi^I_a{\cal D}\sigma{\cal D}\rho] e^{-\frac{1}{g_0}\int L_{HS}}\\
L_{HS}&=&\frac{1}{2}\Phi^I_a(-\partial^2)\Phi^I_a +\frac12\sigma_{ab}(\Phi^I_a\Phi^I_b-\rho_{ab})\nonumber \\
&&+\frac{\mu_0^2}{2}\,tr\rho
- {\det}_2^{(N)}(\rho)
\eeqn
The measure factors are Haar measures for the real symmetric $N\times N$ matrix
fields $\sigma$ and $\rho$. We will analyze this theory in the saddle point approximation at large $N_f$, assuming that the saddle points are homogeneous in space-time. 
Note that the argument of the exponential in the path integral is not
invariant under separate transformations of $\rho$ and $\sigma$
(because of the $Tr\ \sigma\rho$ term). However, the form of the
saddle point equation for $\rho$ is $ \osigma = 2\orho+(\mu_0^2-2(tr\orho))  1$
(where overlines denote saddle point values).  This is a local equation
which has the following important property (of a saddle point): in a
basis where $\orho$ is diagonal, $\osigma$ is diagonal as well.
In this sense, we can simultaneously diagonalize $\sigma_{ab}$ and $\rho_{ab}$, and perform the $\Phi$ path integral. As a result we get 
\begin{eqnarray}
{\cal Z}&=&\int [{\cal D}\sigma{\cal D}\rho] e^{-\frac{N_f}{g_0} S_{\rm eff}(\sigma,\rho)}\,,\\
S_{\rm eff}&=&\frac{1}{2}\sum_{a=1}^N\left(g_0{\rm Tr}\ln(-\partial^2+\sigma_a)-\int \sigma_a\rho_a\right) \nonumber \\
&&+\frac{\mu_0^2}{2}\sum_{a=1}^N\int\rho_a-\sum_{a<b}^N\int \rho_a\rho_b
\label{D3effH-S}\,,
\end{eqnarray}
where $\sigma_a$ and $\rho_a$ are eigenvalues of $\sigma_{ab}$ and $\rho_{ab}$ respectively. To obtain (\ref{D3effH-S}), we rescaled 
$g_0\mapsto g_0/N_f$.  For constant configurations, the effective action yields the effective potential as $V_{\rm eff} =S_{\rm eff}/g_0(Vol_4)$. For large $N_f$, the uniform saddle points $\osigma_a, \orho_a$ satisfy
\beq
\label{SPsigma}
\frac{\delta V_{{\rm eff}}}{\delta\osigma_a}=0 \Rightarrow  \orho_a=g_0\int \frac{d^4p}{(2\pi)^4}\frac{1}{p^2+\osigma_a}\,.
\eeq
Introducing a cutoff $\Lambda$ in equation (\ref{SPsigma}) yields
\beqn\label{SPsigma1}
r_a\equiv \orho_a-\orho_{cr}&=&\frac{g_0}{16\pi^2}\osigma_{a}\ln (\osigma_a/\Lambda^2)+O(\frac{\osigma_a}{\Lambda^2})\,,\\
\orho_{cr}&=&g_0\int^\Lambda \frac{d^4p}{(2\pi)^4}\frac{1}{p^2}=\frac{g_0}{16\pi^2}\Lambda^2\,.
\eeqn
Substituting $r_a$ back into (\ref{D3effH-S}) we first encounter the usual field independent quartic divergence (vacuum energy) which we drop. We also encounter a quadratic divergence which can be cancelled by a fine tuning of the bare mass (in the sense of tuning to a UV fixed point). Namely, we require that 
\beq
\label{Veff_ra}
\frac{\delta V_{\rm eff}}{\delta r_a}=-\frac{1}{2}\osigma_a+\frac{\mu_0^2}{2}-\sum_{b\neq a}(r_b+\orho_{cr}) \,,
\eeq
is finite and zero for $\osigma_a,r_a=0$, which leads to $\mu_0^2=2(N-1)\orho_{cr}$.  Alternatively,  we may renormalize this model by subtracting an infinite contribution to the $\sigma$ tadpole.
Integrating then (\ref{Veff_ra}) and using (\ref{SPsigma1}) we obtain 
\beqn\label{SeffD3fin}
V_{{\rm eff}} & =&  \frac{-1}{64\pi^2}\Biggl[\sum_{a=1}^N\osigma_a^2\ln\frac{e^{1/2}\osigma_a}{\Lambda^2} \nonumber \\
&& +\frac{g_0}{4\pi^2}\sum_{a<b}^N\osigma_a\osigma_b\ln\frac{\osigma_a}{\Lambda^2}\ln\frac{\osigma_b}{\Lambda^2}\Biggl].
\eeqn
To find the saddle point, we should minimize (\ref{SeffD3fin}) with respect to the $\osigma_a$'s. The last term in (\ref{SeffD3fin}) demonstrates the following property of the 4d subdeterminant model. Although the validity of the saddle point is given by large $N_f$, the effective potential is dominated by the sum over the off-diagonal elements of a symmetric matrix, in close proximity with the $D3$-brane potential (\ref{zarembo}) if we take $N$ to be large. Moreover, the $\osigma_a$ are proportional to the (mass)$^2$ of the $N$ vector fields $\Phi^I_a$. 
For simplicity we can consider the homogeneous configuration  $\osigma_a=\sigma$, $a=1,2,..,N$ when the potential becomes in the large-$N$ limit 
\beq\label{VeffR}
V_{{\rm eff}} =-g_0\frac{N^2}{512\pi^4}\sigma^2\ln^2\frac{\sigma}{\Lambda^2}+O(N)\,.
\eeq
\myfig{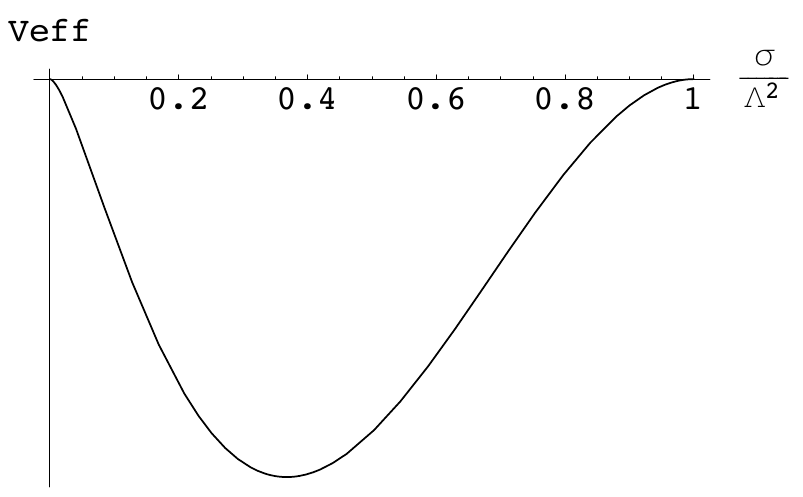}{8}{Plot of $V_{eff}$ for $g_0>0$.}
In Fig.1 we sketch  the effective potential (\ref{VeffR}) for $g_0>0$, which shows the similarity with the corresponding behavior of the effective potential (\ref{zarembo}). However, since the simplicity of our model does not allow to fix either the value or the sign of $g_0$, such a similarity cannot be taken  too far. 
Nevertheless, this is enough motivation to apply the idea to the more intriguing case of $M2$-branes in what follows. 




In $d=3$, the corresponding subdeterminant potential is a special $\varphi^6$ model
\beq
\label{sd3pot}
V_N^{(3)}(\Phi^I_a) =\frac{1}{6(N-3)!}\epsilon^{abcd\cdots}\epsilon^{efg}_{\phantom{egf}d\cdots}\Phi^I_a\Phi^J_b\Phi^K_c\Phi^I_e\Phi^J_f\Phi^K_g\,,
\eeq
where $a,b,..,f=1,2,..,N$, and $I,J,K=1,2,..,N_f$. Notice that (\ref{sd3pot}) is proportional to the BLG scalar potential \cite{bl}  for $N=4$ and $N_f=8$. Hence, (\ref{sd3pot}) may be viewed as a generalization of the BLG potential in the same way that (\ref{sd2pot}) is a generalization of the $SU(2)$ YM potential. Introducing as above the symmetric matrices $\rho_{ab}=\Phi^I_a \Phi^I_b$,
the potential becomes
\beq
\label{ds3potrho}
V_N^{(3)}\equiv \ {\det}_3^{(N)}(\rho)= \frac{1}{6}\left[ (tr\rho)^3-3(tr\rho)(tr\rho^2)+2tr\rho^3\right]\,.
\eeq
Now we introduce the {\it 3d subdeterminant model} with Lagrangian
\beq
\label{L3d}
L_{3d} = \frac{1}{2}\Phi^I_a(-\partial^2)\Phi^I_a +\frac{\mu_0^2}{2}\Phi^I_a\Phi^I_a +g_0V_{N}^{(4)}(\Phi^I_a) -\lambda_0^2V_N^{(3)}(\Phi^I_a)
\eeq
where now $\mu_0^2$, $g_0$ are dimensionful couplings \cite{f1}, while $\lambda_0$ is a dimensionless one. 
As in $d=4$, the scalar theory (\ref{L3d}) at large $N_f$ can be studied using the HS procedure with an auxiliary scalar field $\sigma_{ab}$. After the rescaling $\Phi^I_a\mapsto \Phi^I_a/\sqrt{\lambda_0}$ we can follow the argument above and diagonalize $\rho_{ab}$ and $\sigma_{ab}$ to obtain the effective action
\begin{eqnarray}
{\cal Z}&=&\int [{\cal D}\sigma{\cal D}\rho] e^{-\frac{N_f}{\lambda_0}S_{\rm eff}(\sigma,\rho)}\,,\\
S_{\rm eff}&=&\frac{1}{2}\sum_{a}^N\left(\lambda_0 {\rm Tr}\ln(-\partial^2+\sigma_a)-\int \sigma_a\rho_a\right)\nonumber \\
&& \hspace{-1.4cm}+\frac{\mu_0^2}{2}\sum_{a=1}^N\int \rho_a+\frac{g_0}{\lambda_0}\sum_{a<b}^N\int\rho_a\rho_b -\sum_{a<b<c}^N\int \rho_a\rho_b\rho_c\label{M2effH-S}\,,
\end{eqnarray}
To achieve this form we rescaled $\lambda_0 \mapsto \lambda/ N_f$ and $g_0\mapsto g_0/N_f$.   Following
the same procedure as in the 4-dimensional case, we look for uniform saddle points $\osigma_a$, $\orho_a$ of the large-$N_f$ effective potential $V_{\rm eff} =S_{\rm eff}/\lambda_0 (Vol_3)$ which satisfy
\beq
\frac{\delta V_{\rm eff}}{\delta \osigma_a} =0\Rightarrow \frac{1}{\lambda_0}\orho_a=\int\frac{d^3p}{(2\pi)^3}\frac{1}{p^2+\osigma_a}=\frac{\Lambda}{2\pi^2}-\frac{\osigma_a^{1/2}}{4\pi} +\cdots 
\eeq
Then, writing $r_a\equiv \orho_a-\orho_{cr}\,,\,\,\,\orho_{cr}=\frac{\lambda_0\Lambda}{2\pi^2}$,
we express  the effective potential in terms of the $r_a$ and fine-tune the bare couplings $\mu_0^2$ and $g_0/\lambda_0$ to renormalize it (after dropping as usual the cubic divergence that corresponds to the 3d vacuum energy). In this case we need to impose two renormalization conditions
\beq
\label{renorm3d}
\frac{\delta V_{\rm eff}}{\delta r_a}\Biggl|_{\osigma_a,r_a=0}=0=\frac{\delta ^2 V_{\rm eff}}{\delta r_a\delta r_b}\Biggl|_{r_c=0}\,.
\eeq
We then obtain
$\frac{\mu_0^2}{2}+\left(\!\!\begin{array}{c} N-1\\2\end{array}\!\!\right)\orho_{cr}^2=0$ as well as $\frac{g_0}{\lambda_0}=(N-2)\orho_{cr}$. 
Finally, integrating the first equation of (\ref{renorm3d}) for generic $\osigma_a$ and $\orho_a$ we obtain the effective potential as
\beq\label{ensad}
V_{{\rm eff}}=\frac{1}{24\pi}\left(\sum_{a=1}^{N}\osigma_a^{3/2}
+\frac{6\lambda_0^2}{(4\pi)^2}\sum_{a<b<c}(\osigma_a\osigma_b\osigma_c)^{1/2}\right)\,.
\eeq 

As before, for homogenous configurations $\osigma_a=\sigma$, large-$N$ and assuming that $\lambda_0^2\sim O(1)$ the effective potential becomes
\beq
\label{Veff3d}
V_{\rm eff}=\lambda_0^2\frac{N^3}{384\pi^3}\sigma^{3/2}+O(N^2)\,.
\eeq
The stable vacuum is at $\sigma=0$ \cite{f2}. Notice the peculiar $N^3$ scaling of the effective potential that arises from the three-body nature of the subdeterminant interaction i.e. from the $(N$ choose $3)$ term. We should note that this scaling could have been obtained just from a $(\Phi^I_a\Phi^I_a)^3$ term, which is the large-$N$ limit of the (\ref{sd3pot}) potential. In that sense, although the potential (\ref{sd3pot}) does coincide with the BGL potential for $N=4$,  its algebraic structure does not play a significant role in the large-$N$ result (\ref{Veff3d}).  The latter result is an indication that our model provides the large-$N$ effective description of a system of $M$5-branes. In such a picture 
the three-body interactions would correspond to string junctions \cite{lb}.


We conclude with some observations regarding our subdeterminant potentials and their algebraic properties. Much of the combinatoric structure that we have discussed here arises from properties of the symmetric polynomials involved in the potentials. With this in mind, it is tempting to view our $d=3$ and $d=4$ subdeterminant models as arising from a more general scheme. As an example, note that if we were to identify $\rho$ with some sort of curvature 2-form, then the subdeterminant potentials correspond to their Chern characters.


Next, we comment on an intriguing relationship of our models with spin-systems. In the 4d case we can define the $N\times N$ matrices $T^A$ as
$\epsilon_{ab}^{\phantom{ab}c_1,..c_{N-2}}\equiv \left(T^A\right)_{ab}$,
where we have introduced the collective index $\{c_1,..,c_N\}\mapsto A=1,2,..,\frac{1}{2}N(N-1)$ and $N\geq 3$.
One can show that in fact $\left(T^A\right)_{ab}$ is the fundamental $N$-dimensional representation of $O(N)$, namely $
[T^A,T^B] =f^{AB}_{\phantom{AB}C}T^C$.
Hence, the potential (\ref{sd2pot}) can be written as
\begin{eqnarray}
V_N^{(4)} &=&\frac{1}{2(N-2)!}\delta_{AB}\Bigl(\Phi^I_a\left(T^A\right)_{ab}\Phi^J_b\Bigl) \Bigl(\Phi^I_e\left(T^B\right)_{ef}\Phi^J_f\Bigl) \nonumber \\
\label{ds2potON}
&=&\frac{1}{2(N-2)!}\delta_{AB}(S^A)^{IJ}(S^B)^{IJ}\,,
\end{eqnarray}
 with the obvious identifications. 
We can think of the $(S^A)^{IJ}$ as classical $O(N)$ spins that carry the antisymmetric indices $IJ$. It is necessary that $N_f\geq 2$ for the potential to be non-trivial. 
It is furthermore interesting to note that the $(S^A)^{IJ}$ can be elevated to {\it quantum} $O(N)$ spins obeying $[S^A,S^B]=f^{AB}_{\phantom{AB}C}S^C$ if the $\Phi^I_a$'s are promoted to operators with non-trivial commutation relations. 
Consider $N_f=2$, namely $I,J=1,2$. Then there is only one independent $(S^A)^{IJ}$ in (8), namely,
$\left(S^A\right)^{IJ}\mapsto S^A =\phi^1_a\epsilon^A_{\phantom{A}ab}\phi^2_b$.
Now we can compute the commutator
$[S^A,S^B] = \left[\epsilon^A_{\phantom{A}ab}\,\epsilon^B_{\phantom{A}cd}-\epsilon^B_{\phantom{A}ab}\,\epsilon^A_{\phantom{A}cd}\right]\phi^1_a\phi^2_b\phi^1_c\phi^2_d
$.
By imposing the 
commutation relations
$[\phi^1_a,\phi^2_b]=\delta_{ab}, [\phi^1_a,\phi^1_b]=[\phi^2_a,\phi^2_b]=0$
which is equivalent to a set of  $N$ Heisenberg algebras,
we obtain
$[S^A,S^B]= f^{AB}_{\phantom{AB}C}S^C$.
Therefore the $S^A$ are a representation of $O(N)$. For $N=3$, the above is the usual Schwinger boson representation of $O(3)$ \cite{fradkin}. 

Now, it is tempting to generalize this construction to the 3d case. We define the {\it cubic matrices} $T^A$ as 
$\epsilon_{abc}^{\phantom{abc}c_1..c_{N-3}}=\left(T^A\right)_{abc}$
where $A\equiv\{ c_1..c_{N-3}\}=1,2,..,\frac{1}{3!}N(N-1)(N-2)$ and $N\geq 4$.
Given that, we can express  the 3d subdeterminant potential in terms of ``generalized spins" $(S^A)^{IJK}$ as
\begin{eqnarray}
V_N^{(3)} &=& \frac{\delta_{AB}}{6!(N-3)!}\left[\left(T^A\right)_{abc}\Phi^I_a\Phi^J_b\Phi^K_c\right] \left[\left(T^B\right)_{efg}\Phi^I_e\Phi^J_f\Phi^K_g\right]\nonumber \\
\label{sd3potnew} &=& \delta_{AB}(S^A)^{IJK}(S^B)^{IJK}\,.
\end{eqnarray}
For generic $N$ and $N_f$ one should be able to study the algebraic structure of the {\it cubic matrices} $T^A$ \cite{cubic,Ho,Li} as well as of the ``generalized spins" $(S^A)^{IJK}$ \cite{lmmp}. However, for the minimal case  $N=4$ and $N_f=3$, the cubic matrices $T^A$ become the usual 4-indexed Levi-Civita tensors 
$(T^A)_{abc}\equiv (\epsilon^A)_{abc}$ and also
$(S^A)^{IJK} \mapsto S^A=(\epsilon^A)_{abc}\Phi^1_a\Phi^2_b\Phi^3_c$.
In this case, one can define the cubic matrices \cite{cubic,Ho,Li} $
\left(T^A\right)_{abc} =\left|\epsilon^A_{\phantom{A}abc}\right|e^{\frac{{\rm i}\pi}{8}\epsilon_{Aabc}}$
and the following multiplication rule 
\beq\label{cubicmatrix}
\sum_{m,n,k}\left(T^A\right)_{abm}\left(T^B\right)_{anc}\left(T^C\right)_{kbc}\,\Delta_{mnk} =\left(T^AT^BT^C\right)_{abc}\,.
 \eeq
We have introduced the generalized Kronecker $\Delta_{abc}$ which is 1 for $a=b=c$ and zero otherwise. All indices run from $1,...,4$, but {\it no summation} over repeated indices is implied.   
With the above definitions one can show that the standard ternary commutator satisfies
\beq
\label{ternary}
[T^A,T^B,T^C]_{abc}=-{\rm i}\epsilon^{ABC}_{\phantom{ABC}D}\left(T^D\right)_{abc}\,,
\eeq
i.e. $(T^A)_{abc}$ are a representation of the ${\cal A}_4$ 3-algebra \cite{bl,cubic,Ho,Li}. 

It would be natural then to study the ternary commutator of the ``spins" $S^A$ when the $\Phi^I$'s are promoted into generalized Nambu-Heisenberg oscillators as in \cite{tak}. This way we expect to obtain a generalization of the Hubbard model for spins that satisfy the ${\cal A}_4$ 3-algebra \cite{lmmp}. 
 
In conclusion, we have initiated the study of the 4d and 3d subdeterminant models. These models can be studied using a Hubbard-Stratonovich transformation and exhibit quite interesting large-$N$ behavior which seems intimately connected with the behavior of the effective theory describing the stringy interactions among $D3$-branes in 4d. Hence, it is conceivable that the 3d subdeterminant model might be related to the effective theory that describes the interactions among $M2$-branes. The $N^3$ scaling of the 3d effective potential would then imply that this effective theory is related to $M5$-branes. 
Finally, we have pointed out the versatility of our subdeterminant potentials, which in 4d provide the $O(N)$ generalization of Schwinger bosons, while in 3d give a new ``generalized spin" model.
In conclusion, we note that we have also done a preliminary
study of the supersymmetric extension of our $d=3$ model, its relation to
the $d=3$ SYM theory (following \cite{mp}) and the appearance of new algebraic
structures \cite{f3}. These results will be presented 
in detail elsewhere \cite{lmmp}.

{\bf Acknowledgments:} 
\small
We wish to thank P. Argyres, H. Awata, O. Bergman, J. Gomis, 
M. Henningson,
N. Lambert, A. Tseytlin, V. Scarola, G. Siopsis, G. Semenoff and E. Sharpe for interesting conversations concerning this work.
DM is supported in part by the U.S.\ DOE under contract DE-FG05-92ER40677.
ACP is partially supported by the FP7-REGPOT-2008-1 grant CreteHEPCosmo No 28644  and RGL receives partial 
support from the U.S. DOE, under contract DE-FG02-91ER40709.
This work was initiated and later developed in the respective stimulating atmospheres of the University of Crete and the Aspen Center for Physics.


\begin{thebibliography}{99}

\bibitem{Zarembo}
K.~Zarembo,
  Phys.\ Lett.\  B {\bf 462}, 70 (1999)
  A.~A.~Tseytlin and K.~Zarembo,
  Phys.\ Lett.\  B {\bf 457} 77 (1999).
  
\bibitem{bl} 
 J.~Bagger and N.~Lambert,
  Phys.\ Rev.\  D {\bf 75}, 045020 (2007)
  Phys.\ Rev.\  D {\bf 77}, 065008 (2008)
  JHEP {\bf 0802}, 105 (2008)
A. ~Gustavsson, Nucl. \ Phys. \ B {\bf 811} 66 (2009)
  and references therein.
  
  
\bibitem{fradkin}
For an insightful review consult, E.~Fradkin, {\it Field Theories of Condensed Matter Systems, } Frontiers of Physics,
Perseus Books, 1991 and references therein.

\bibitem{f1}
Again, we could describe the  renormalization of the saddle point equations  by simply subtracting the $\sigma$ tadpole. This is equivalent to the set of relevant counterterms shown here.

\bibitem{f2}
Had we started with a plus sign in the last term in (\ref{L3d}) we would have obtained a minus sign in the sum in (\ref{ensad}). In this case the effective potential would have a flat direction when the bare coupling reaches a critical value $\lambda_{cr}^2N(N-1)=16\pi^2$. The potential becomes unbounded from below for $\lambda_0^2>\lambda_{cr}^2$. See,
W.~A.~Bardeen, M.~Moshe and M.~Bander,
  Phys.\ Rev.\ Lett.\  {\bf 52} (1984) 1188.

  
\bibitem{lb}
D.~Berenstein and R.~G.~Leigh,
  Phys.\ Rev.\  D {\bf 60}, 026005 (1999).
 

\bibitem{lmmp}
R.~G.~Leigh, A.~Mauri, D.~Minic and A.~C.~Petkou, in preparation.

\bibitem{cubic}
Y. Kawamura, Prog.Theor.Phys. {\bf 110} 579 (2003) . 

\bibitem{Ho}
 P.~M.~Ho, R.~C.~Hou and Y.~Matsuo,
  JHEP {\bf 0806} (2008) 020
  [arXiv:0804.2110 [hep-th]].

\bibitem{Li}
 M.~Li and T.~Wang,
  JHEP {\bf 0807} (2008) 093
  [arXiv:0805.3427 [hep-th]].

\bibitem{tak}
 R.~Chatterjee and L.~Takhtajan,
  Lett.\ Math.\ Phys.\  {\bf 37}, 475 (1996).
  
  
\bibitem{mp}
A.~Mauri and A.~C.~Petkou,
  Phys.\ Lett.\  B {\bf 666}, 527 (2008).
  
\bibitem{f3}
H.~Awata, M.~Li, D.~Minic and T.~Yoneya,
  JHEP {\bf 0102}, 013 (2001).


\end{thebibliography}
\end{document}